\documentclass[conference]{IEEEtran}
\IEEEoverridecommandlockouts
\usepackage{cite}
\usepackage{amsmath,amssymb,amsfonts}
\usepackage{algorithmic}
\usepackage{graphicx}
\usepackage{textcomp}
\usepackage{xcolor}
\usepackage{multirow}
\usepackage{array}     
\def\BibTeX{{\rm B\kern-.05em{\sc i\kern-.025em b}\kern-.08em
    T\kern-.1667em\lower.7ex\hbox{E}\kern-.125emX}}
\begin{document}

\title{Experimental Evaluation of a Low-Power Ultra-Wideband Receiver for Spectrum Sensing\\}

\author{
    \IEEEauthorblockN{Panagiotis Vlachos\IEEEauthorrefmark{2}, Ioannis A. Bartsiokas\IEEEauthorrefmark{1}, Cedric Dehos\IEEEauthorrefmark{3}, Francois Rivet\IEEEauthorrefmark{2}, George Karachalios\IEEEauthorrefmark{1}}
    \IEEEauthorblockA{\IEEEauthorrefmark{1}\textit{Intracom Defense, Koropi, 19441, Greece}}
    \IEEEauthorblockA{\IEEEauthorrefmark{2}\textit{IMS Laboratory, University of Bordeaux, CNRS, Bordeaux INP, IMS, UMR 5218, F-33400 Talence, France}}
    \IEEEauthorblockA{\IEEEauthorrefmark{3}\textit{CEA-Leti, Université Grenoble-Alpes, Minatec Campus Grenoble, France}}
}

\maketitle

\begin{abstract}
Walsh-sequence-based receiver architectures offer an alternative approach for the reception and reconstruction of multiple simultaneous RF signals over wide bandwidths. While previous works have focused on the architecture and theoretical operation of Walsh-domain processing, limited experimental results have been reported so far on hardware operation under realistic conditions. This paper presents the experimental evaluation of a Walsh-sequence-based receiver prototype. A measurement campaign was conducted under various signal placement, interference, and receiver configuration scenarios. Reconstructed spectra, Walsh-lane power spectral densities (PSDs), and Error Vector Magnitude (EVM) measurements were used to investigate Walsh-domain signal representation, interference effects, and receiver performance. Finally, dense- spectrum experiments involving up to ten simultaneously active signals demonstrate the potential of the architecture for low power wideband spectrum sensing applications.
\end{abstract}


\section{Introduction}
Future wireless communication systems, including sixth-generation (6G) networks, are expected to operate in increasingly congested spectral environments where multiple communication signals coexist with strong interferers within a limited bandwidth \cite{b1}. Joint Communication and Sensing (JCAS) is also expected to become a key feature of next-generation networks \cite{b2}. These trends have motivated significant research into wideband spectrum sensing and acquisition architectures capable of reducing digitization complexity while maintaining the ability to process multiple signals over large bandwidths \cite{b3,b4new}. Existing approaches include channelized receivers, compressed sensing techniques, and sub-Nyquist sampling architectures that exploit signal sparsity to relax conventional sampling requirements. Among these approaches, Walsh-sequence-based architectures have recently emerged as a promising solution for wideband signal acquisition and multi-signal reception. Their application has been investigated in arbitrary waveform generation \cite{b5}, digital predistortion \cite{b6}, wideband channel aggregation \cite{b7}, massively parallelized receiver implementations \cite{b8}, and spectrum sensing applications \cite{b9}. However, experimental investigations of Walsh receivers remain limited. While \cite{b8} introduced the architecture, this work provides the first extensive experimental characterization under realistic interference-rich conditions. Specifically, the paper investigates: Walsh-domain signal representation and interference suppression through adaptive transconductance control, the impact of ADC-offset-induced spectral images on reconstruction quality, nonlinear distortion mechanisms generated by multiple strong interferers, and the capability of the receiver to identify occupied channels under dense-spectrum conditions containing up to ten simultaneously active signals. These measurements provide practical insight into the strengths and limitations of Walsh-domain processing for future wideband sensing and communication systems. 

\section{Receiver Architecture and Experimental Setup}
\begin{figure}[t]
\centerline{\includegraphics[width=0.5\textwidth]{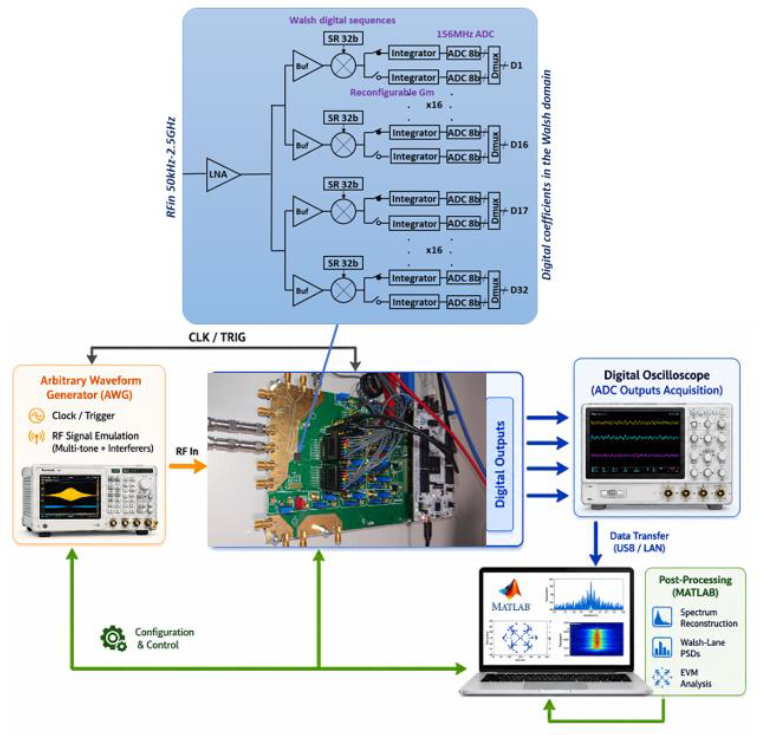}}
\caption{Simplified Walsh receiver architecture (top) and experimental setup for Walsh-domain signal acquisition, reconstruction, and performance evaluation (bottom).}
\label{fig2}
\end{figure}

The experimental setup used in this work is shown in Fig. \ref{fig2}. The test signals and interference are generated offline in MATLAB and uploaded to an Arbitrary Waveform Generator (AWG). The generated waveform is applied to the receiver prototype, where signal projection and reconstruction is performed. The received IF signal, operating over a frequency range from 50 kHz to 2.5 GHz, is processed through 32 parallel Walsh branches. Each branch produces Walsh-domain coefficients. For that purpose, the signal is projected onto a set of orthogonal sequences, where each branch performs a multiplication with the corresponding Walsh sequence followed by integration and analog-to-digital conversion. The resulting Walsh coefficients provide an alternative representation of the received spectrum in the Walsh domain. Signal reconstruction is performed digitally by combining the coefficients according to the selected receiver configuration. Each lane includes a digital multiplier, programmable integration stage, and time-interleaved SAR ADCs operating at a nominal output rate of 156.25 MSps. The architecture supports programmable Walsh sequences, branch-dependent transconductance control $G_m$, adjustable integration windows, and adaptive receiver reconfiguration. Additional implementation details of the receiver architecture are provided in \cite{b8}.

The digital receiver outputs were captured using a high-speed oscilloscope and transferred to a computer for offline processing. Reconstructed spectra, Walsh-lane power spectral densities (PSDs), and Error Vector Magnitude (EVM) measurements were used to evaluate receiver performance under different signal placement, interference, and gain control configurations.

\section{Performance Evaluation and Experimental Characterization}

A measurement campaign was conducted to evaluate the behavior of the Walsh receiver under different signals and interference conditions.

\subsection{Walsh-Domain Reconstruction and Spectral Images}

A first measurement was performed using a desired 50 MHz 16-QAM signal centered at 1.10 GHz and a 20 MHz interferer at 1.25 GHz with a relative power level of +10 dB. Unless otherwise specified, all Walsh branches were configured with $G_m = 30$, while selected branches used for interference mitigation were set to $G_m = 1$.

\begin{figure}[t]
\centerline{\includegraphics[width=0.8\linewidth]{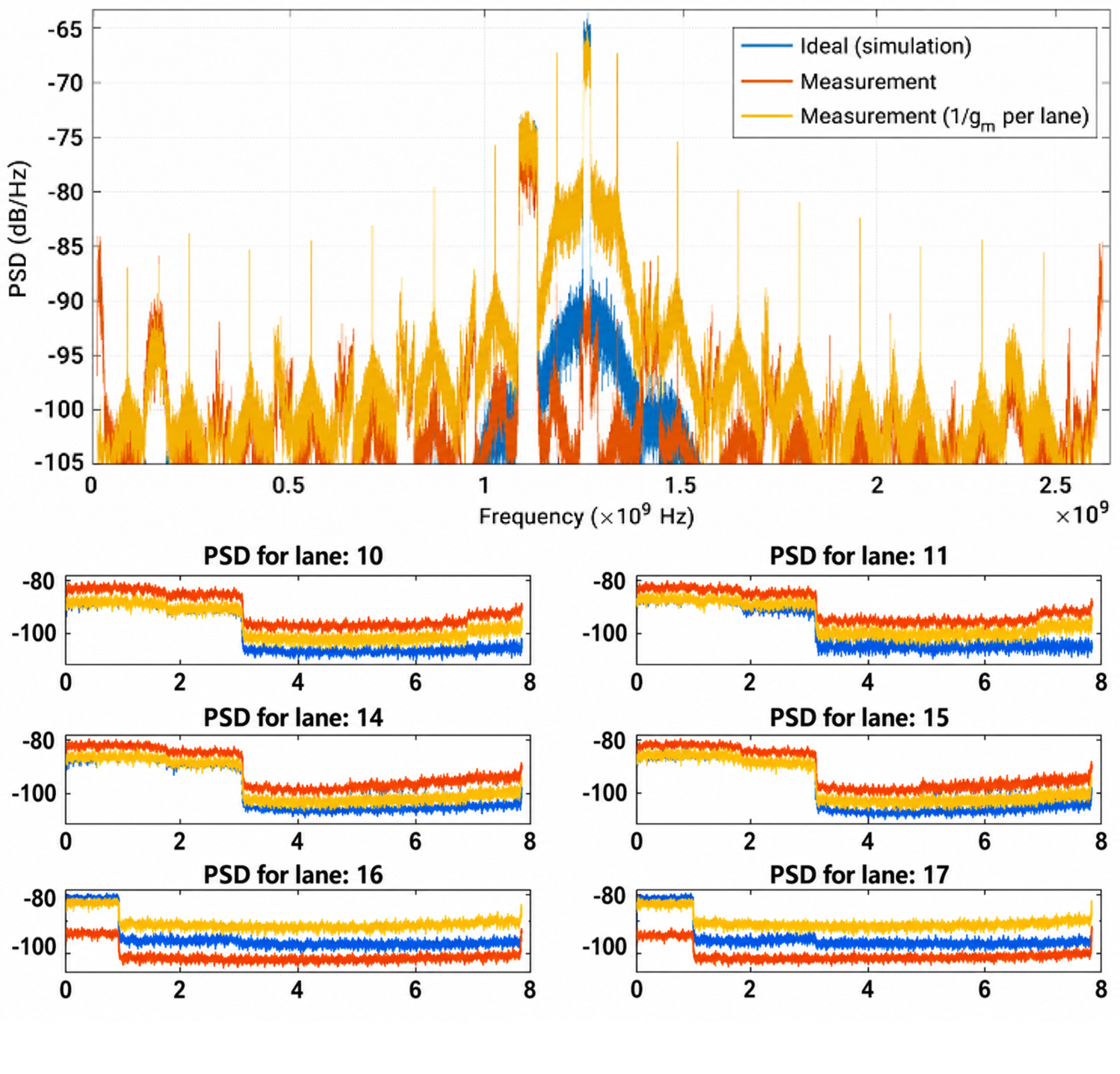}}
\caption{Walsh receiver response for a desired signal at 1.10 GHz and interferer at 1.25 GHz. (a) Reconstructed spectrum. (b) PSDs of the dominant Walsh lanes}
\label{fig3}
\end{figure}

\begin{figure}[t]
\centerline{\includegraphics[width=0.8\linewidth]{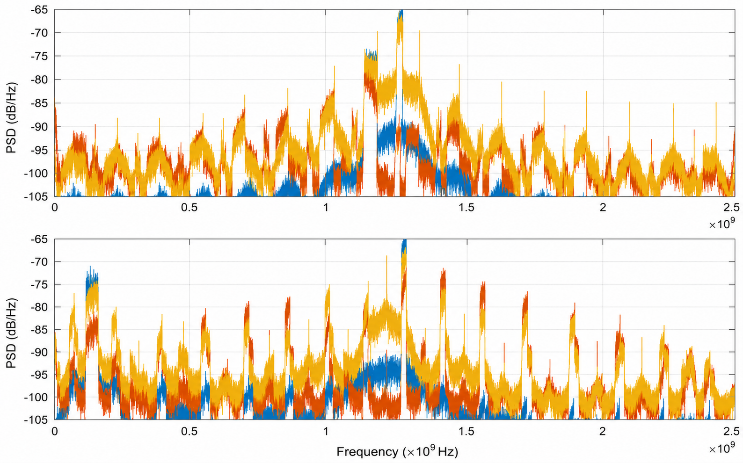}}
\caption{Reconstructed spectra for (a) Desired signal at 1.15 GHz with an interferer at 1.25 GHz. (b) Desired signal at 150 MHz with an interferer at 1.32 GHz}
\label{fig5}
\end{figure}

The objective was to investigate Walsh-domain signal representation and the effectiveness of branch-dependent gain control. The reconstructed spectrum and corresponding Walsh-lane PSDs are shown in Fig. \ref{fig3}. Additional measurements were performed with the desired signal centered at 1.15 GHz and 150 MHz, with interferers at 1.25 GHz and 1.32 GHz, respectively. The reconstructed spectra are shown in Fig. \ref{fig5}, yielding EVM values of 13.89\%, 36.23\%, and 48\%. The main observations are:
\begin{itemize}
    \item \underline{Walsh-Domain Signal Representation}: The desired signal is primarily represented by Walsh lanes 10, 11, 14, and 15, while lanes 16 and 17 are dominated by the interferer. Due to its central position in the band, the desired signal energy is distributed across multiple Walsh coefficient pairs (10–11 and 14–15), corresponding to the same frequency region with different Walsh-code phases.
    \item \underline{Interference Mitigation}: Reducing the transconductance of Walsh lanes 16 and 17 to $G_m = 1$ significantly attenuates the blocker contribution while preserving the desired signal.
    \item \underline{ADC-Offset-Induced Spectral Images}: Signal placement near 78.125 MHz and multiples of 156.25 MHz produces spectral images and spurious components caused by periodic ADC offset components introduced by the acquisition scheme.
    \item \underline{Limitation of Walsh-Domain Gain Control}: Although attenuating the Walsh branches associated with the interferer reduces its direct contribution, it does not suppress the spectral images limitation.
\end{itemize}

These results demonstrate that adaptive Walsh-domain gain control is effective for suppressing interferers. However, ADC-offset-induced spectral images remain a significant performance limitation.
    
\subsection{Intermodulation in the Presence of Multiple Interferers}

To further investigate nonlinear distortion mechanisms, a second set of measurements was performed using two strong interferers located at 625 MHz and 1.25 GHz. The objective was to evaluate the effectiveness of Walsh-domain gain control in the presence of multiple interference sources.

\begin{figure}[b]
\centerline{\includegraphics[width=0.8\linewidth]{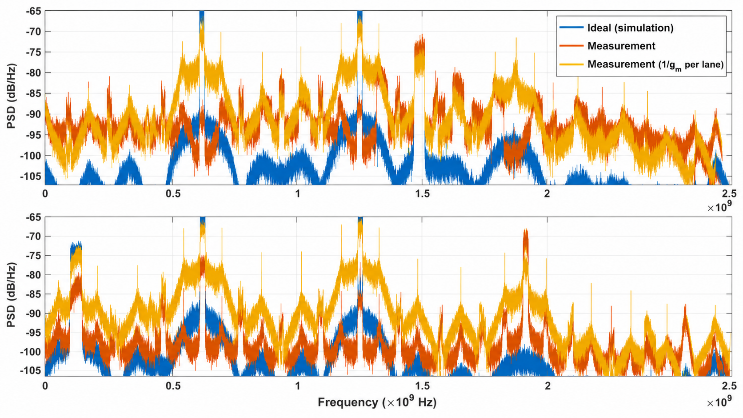}}
\caption{Reconstructed spectrum with interferers located at 625 MHz and 1.25 GHz for (a) 1.5 GHz (b) 150 MHz}
\label{fig6}
\end{figure}

Two measurements were conducted with the desired signal centered at 1.5 GHz and 150 MHz, respectively, while maintaining the same interferer configuration. The corresponding reconstructed spectra are shown in Fig. \ref{fig6}. To mitigate the interference contribution, the transconductance of selected Walsh branches associated with the dominant interferers and intermodulation products was reduced to $G_m = 1$, while all remaining branches were maintained at their nominal value. For the 1.5 GHz scenario, Walsh lanes 8, 9, 16, 17, 24, and 25 were attenuated, whereas for the 150 MHz scenario, Walsh lanes 8, 9, 16, and 17 were attenuated. The main observations are:
\begin{itemize}
    \item \underline{Interference Attenuation}: Reducing the gain of the selected Walsh branches significantly suppresses the dominant interferer contributions in both scenarios.
    \item \underline{Persistent Intermodulation Distortion}: Despite the applied gain control, several spurious spectral components remain visible throughout the reconstructed spectrum, indicating that nonlinear distortion products are distributed across multiple Walsh coefficients.
    \item \underline{Reconstruction Quality}: The reconstructed signals exhibit EVM values of 33.91\% and 43.65\% for the 1.5 GHz and 150 MHz desired-signal cases, respectively, representing a substantial degradation compared to the single-interferer scenarios.
    \item \underline{Limitation of Walsh-Domain Gain Control}: While adaptive Gm control effectively reduces the direct contribution of strong interferers, it cannot completely suppress the intermodulation products generated under multi-interferer conditions.
\end{itemize}

These results demonstrate that gain control remains effective for mitigation of multiple blockers. However, intermodulation products is the dominant performance limitation of the current receiver implementation. Nonlinear products affect multiple Walsh coefficients that cannot be mitigated through branch-level gain adaptation. 

\subsection{Dense Spectrum Reconstruction and Spectrum Sensing}
To evaluate the suitability of the Walsh receiver for spectrum sensing applications, the architecture was tested under highly congested spectral conditions. Two scenarios were considered: a medium-density environment containing six desired signals and two strong interferers, and a high-density environment containing ten simultaneously active signals distributed across the analyzed bandwidth.

\begin{figure}[t]
\centerline{\includegraphics[width=0.8\linewidth]{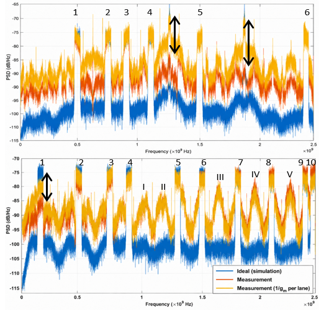}}
\caption{Reconstructed spectra for (a) 6 channels with 2 interferers and (b) 10 channels.}
\label{fig7}
\end{figure}

The reconstructed spectrum for the first scenario is shown in Fig. \ref{fig7}. Based on the observations of Section B, Walsh lanes 8, 9, 16, 17, 24, and 25 were configured with $G_m = 1$, while all remaining branches were maintained at their nominal value. Lanes 16, 17, 24, and 25 are associated with the two strong interferers, whereas lanes 8 and 9 correspond to the dominant intermodulation product generated by the frequency difference between the interferers. The main observations are:
\begin{itemize}
    \item \underline{Interference Mitigation}: The applied gain control reduces the contribution of both the interferers and their dominant intermodulation product.
    \item \underline{Spectrum Identification}: Despite the high spectral occupancy, the receiver successfully identifies the occupied channels and reconstructs their spectral locations.
    \item \underline{Reconstruction Quality}: The measured EVM values for the six reconstructed channels are 26.48\%, 31.41\%, 29.30\%, 19.38\%, 27.54\%, and 21.21\%, respectively.
    \item \underline{Spectrum Sensing Capability}: Reliable signal reconstruction remains possible under moderate spectral congestion, demonstrating the suitability of the architecture for multi-signal sensing scenarios.
\end{itemize}

The reconstructed spectrum for the ten-signal scenario is presented in Fig. \ref{fig7}. In this case, all channels were simultaneously active across the analyzed bandwidth, creating an extreme spectrum occupancy condition. The main observations are:
\begin{itemize}
    \item \underline{Occupancy Detection}: The receiver successfully identifies all occupied frequency regions across the analyzed bandwidth corresponding to a channel occupancy detection rate of 100\%. Five additional artifact-induced peaks appear in unoccupied regions, resulting in a non-zero false-alarm probability.
    \item \underline{Power Redistribution}: The lower-frequency channels exhibit a noticeable attenuation, indicating that part of the signal energy is distributed among higher-order Walsh coefficients.
    \item \underline{Spectral Images}: Several spectral replicas appear in frequency regions where no signals are present, revealing reconstruction artifacts and residual distortion products.
    \item \underline{Reconstruction Quality}: The measured EVM values for the ten reconstructed channels are 49.96\%, 21.21\%, 27.62\%, 28.69\%, 29.68\%, 36.51\%, 41.63\%, 39.31\%, 36.64\%, and 19.42\%, respectively.
\end{itemize}

These results demonstrate that the Walsh receiver can successfully identify occupied channels even under highly congested spectral conditions. Table \ref{tab:results} summarizes the experimental results obtained in this work. 

\begin{table}[!t]
\centering
\caption{Summary of Experimental Results}
\label{tab:results}
\renewcommand{\arraystretch}{1.3}
\setlength{\tabcolsep}{4pt}
\begin{tabular}{p{1.9cm}|c|>{\centering\arraybackslash}p{1.6cm}|>{\centering\arraybackslash}p{2.2cm}}
\hline
\multirow{2}{*}{\textbf{Scenario}} & \multicolumn{3}{c}{\textbf{Results}} \\
\cline{2-4}
 & \textbf{EVM (\%)} & \textbf{Dominant Limitation} & \textbf{Main Observation} \\
\hline
1 Interferer\newline 1.1 GHz Signal & 13.89 & Blocker & Successful suppression \\
\hline
1 Interferer\newline 1.15 GHz Signal & 36.23 & ADC Offset IMD & Spectral images \\
\hline
1 Interferer\newline 0.15 GHz Signal & 48.00 & ADC Offset IMD & Spectral images \\
\hline
2 Interferers\newline 1.5 GHz Signal & 33.91 & IMD & Residual distortion \\
\hline
2 Interferers\newline 0.15 GHz Signal & 43.65 & IMD & Distortion dominates \\
\hline
2 Interferers\newline 6 channels & 19--31 & Congestion & Reliable sensing \\
\hline
10 channels & 19--50 & Congestion + IMD & Occupancy detected \\
\hline
\end{tabular}
\end{table}

\subsection{Power Consumption}

The total receiver power consumption was measured at 46 mW under a nominal supply voltage of 0.8 V, corresponding to a supply current of 57 mA. The measurement confirms the low-power operation of the architecture while supporting wideband reception over a 50 kHz–2.5 GHz bandwidth with 8-bit conversion.

\section{Conclusions}

This paper presented an experimental evaluation of a Walsh-sequence-based receiver under realistic interference-rich conditions. The results demonstrated that adaptive transconductance control can identify and attenuate interferer-dominated Walsh coefficients, improving signal reconstruction and interference mitigation. The measurements also revealed that ADC-offset-induced spectral images and multi-interferer intermodulation products are the dominant limitations of the current implementation, to be corrected in a future chip version.

Dense-spectrum experiments with up to ten simultaneously active signals confirmed that the architecture identifies occupied channels across a 50 kHz–2.5 GHz bandwidth despite ADC-offset-induced spectral images and intermodulation distortion. At a measured 46 mW, it offers an attractive bandwidth-to-power tradeoff for future wideband spectrum sensing.

Future work will focus on AI-assisted adaptive receiver reconfiguration, calibration, and distortion mitigation techniques to further enhance reconstruction accuracy and sensing performance in highly dynamic spectral environments. Furthermore, the wide operating bandwidth and low-power operation of the proposed receiver make it a promising candidate for both current communication systems and future B5G/6G networks, where flexible wideband spectrum awareness, dynamic spectrum access, and adaptive baseband processing capabilities will be essential.

\section*{Acknowledgment}

This work is supported by the European Union’s Horizon 2020 Hermes project under grant agreement N° 964246.

\vspace{12pt}
\color{red}

\end{document}